\documentclass[aps,prl,floatfix,nofootinbib,superscriptaddress,preprintnumbers,amsmath,amssymb,footinbib,twocolumn]{revtex4-1}
\usepackage{textcomp}
\usepackage{mathtools}
\usepackage{amsmath}
\usepackage{amssymb}
\usepackage{esint}
\usepackage{comment}
\usepackage[utf8]{inputenc}
\usepackage{hyperref}
\usepackage[final]{graphics}
\usepackage{amsfonts}
\usepackage{color}
\usepackage{MnSymbol}

\renewcommand{\v}[1]{ \ensuremath{ {\mathbf{#1}_\perp} }}

\begin{document}

%\title{
%Imprints of gluon saturation in small systems
%}
\title{Systematics of azimuthal anisotropy harmonics in proton-nucleus collisions at the LHC from the Color Glass Condensate}

\author{Mark Mace}
\affiliation{Physics Department, Brookhaven National Laboratory, Upton, New York 11973-5000, USA}
\affiliation{Department of Physics and Astronomy, Stony Brook University, Stony Brook, NY 11794, USA}
\author{Vladimir V. Skokov}
\affiliation{RIKEN-BNL Research Center, Brookhaven National Laboratory, Upton, New York 11973-5000, USA}
\author{Prithwish Tribedy}
\affiliation{Physics Department, Brookhaven National Laboratory, Upton, New York 11973-5000, USA}
\author{Raju Venugopalan}
\affiliation{Physics Department, Brookhaven National Laboratory, Upton, New York 11973-5000, USA}

\date{\today}
\begin{abstract}
Simple power counting arguments in the dilute-dense framework of the Color Glass Condensate (CGC) Effective Field Theory predict that 
even and odd azimuthal anisotropy harmonics of two-particle correlations in proton-nucleus collisions at the LHC will respectively satisfy  $v^2_{2n}\{2\} \propto N_{\rm ch}^0$ and $v^2_{2n+1}\{2\} \propto N_{\rm ch}$, where $N_{\rm ch}$ denotes the number of charged particles. % measured. 
We show that these expectations are borne out qualitatively, and even quantitatively, within systematic uncertainties, for $v_2$ and $v_4$ in comparisons with data from the ATLAS collaboration. We also observe that ATLAS data for the $v_3$ azimuthal harmonic are in excellent agreement with our qualitative expectation; quantitative comparisons are numerically challenging at present. The lessons from this study fully complement those gained by the recent comparison of the CGC dilute-dense framework~\cite{Mace:2018vwq} to data from the PHENIX collaboration on small system collisions at RHIC.  
%
%
%Recent measurement allow us to study systematics of the long-range rapidity
%correlations at LHC energies in p-A collisions.  In particular ATLAS demonstrated
%that $v^2_2\{2\}$ is approximately independent of the number of produced charged particles, $N_{\rm ch}$,
%while $v^2_3\{2\}$-dependence on $N_{\rm ch}$ is rather strong. We show that this
%can be naturally explained in the CGC/saturation framework, where $v^2_{2n}\{2\} \propto N_{\rm ch}^0$
%and $v^2_{2n+1}\{2\} \propto N_{\rm ch}$. 
\end{abstract}

\maketitle

In a recent preprint~\cite{Mace:2018vwq}, we showed that the dilute-dense framework of the Color Glass Condensate (CGC) Effective Field Theory (EFT) ~\cite{Gelis:2010nm} 
qualitatively describes the hierarchy of $v_{2,3}(p_\perp)$ azimuthal Fourier harmonic coefficients of rapidity separated two-particle ``ridge" correlations measured by the PHENIX experiment in collisions of proton/deuterium/helium-3 ions off gold ions at center-of-mass energy of $\sqrt{s}=200$ GeV/nucleon~\cite{Aidala:2018mcw}. Within theoretical uncertainties, the CGC EFT computations also provide semi-quantitative agreement with the PHENIX measurements. 

The model to data comparison suggests that initial state correlations in the hadron wavefunctions provide a competitive alternative explanation to models that describe the data in terms of hydrodynamic collectivity of the quark-gluon matter produced in the collisions. This conclusion is fortified by the fact that the systematics of m-particle $v_n\{m\}$ harmonics, previously believed to provide an unambiguous signature of hydrodynamic collectivity, are also reproduced in a simple initial state parton model~\cite{Dusling:2017dqg,Dusling:2017aot,Dusling:2018hsg}.

While theory comparisons to measurements across system size are very important for understanding the underlying physical origin of the ridge two-particle correlations, they are at present available only for events characterized by a limited range in $N_{\rm ch}$, the number of charged particles produced. However extensive data on the $N_{\rm ch}$ dependence of ridge yields (and the corresponding $v_n$ coefficients) is available in proton-lead (p+A) collisions at the LHC at $\sqrt{s}=5.02$ GeV/nucleon. In addition, p+A collisions have the virtue that modeling the proton wavefunction at high energies is simpler than that of the deuteron or helium-3. Reproducing the systematics of the  $v_n$ dependence on $N_{\rm ch}$ in small systems even qualitatively is a challenge for all theory frameworks and can help distinguish between them. 

In this note, we will examine the $N_{\rm ch}$ dependence of the azimuthal Fourier harmonics as measured by the ATLAS experiment in p+A collisions at $5.02$ TeV/nucleon. We will show that the qualitative features of the data can be deduced very simply from the corresponding equations in the dilute-dense approximation of the CGC. We will go one step further and show that the magnitude of $v_2$ and $v_4$, as a function of $N_{\rm ch}$, is reproduced  in the CGC EFT within theoretical uncertainties. For reasons we shall discuss, quantitative results for the $N_{\rm ch}$ dependence of $v_3$ are more challenging numerically -- it is outside the scope of the present work. 

The dilute-dense approximation~\cite{Kovchegov:1998bi,Dumitru:2001ux,Blaizot:2004wu,Kovner:2012jm,Kovchegov:2012nd}  of the CGC EFT consists of keeping terms in the solution of the QCD Yang-Mills equations to compute inclusive gluon amplitudes that are to lowest order in the ratio $\rho_p/k_{\perp,p}^2$ in the projectile but to all orders in the  ratio $\rho_t/k_{\perp,t}^2$ in the target~\footnote{A further glasma graph approximation corresponds to the regime where one expands to lowest order in both $\rho_p/k_{\perp,p}^2$ and $\rho_t/k_{\perp,t}^2$~\cite{Dumitru:2008wn,Dumitru:2010iy,Dusling:2012cg,Dusling:2012wy,Dusling:2013qoz,Dusling:2015rja}.}. Here $\rho_p (\rho_t)$ is color charge density in the proton (lead nucleus) and $k_{\perp,p}$ ($k_{\perp,t}$) is the transverse momentum of the scattered gluon from the proton (lead nucleus). 
While this dilute-dense approximation may be sufficient to compute the even harmonics $v_{2n}$, an accidental parity symmetry sets $v_3=0$ at this order.  This is well known to be an artifact of the leading order in $\rho_p/k_{\perp,p}^2$ approximation. For instance, numerical work in the dense-dense limit of the CGC EFT, where all orders in both $\rho_p/k_{\perp,p}^2$ and $\rho_t/k_{\perp,t}^2$ are kept, clearly recover finite values of $v_3$~~\cite{Lappi:2009xa,Schenke:2015aqa}. 

\begin{figure*}
\includegraphics[width=0.38\linewidth]{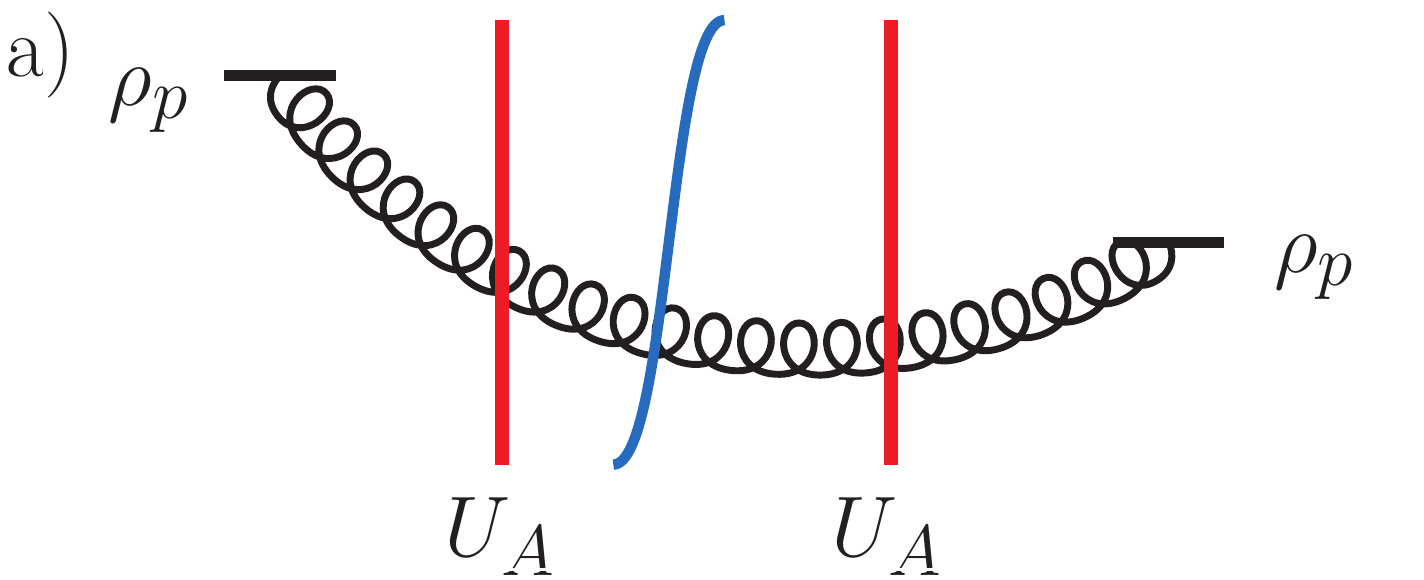}\hspace{1cm}
\includegraphics[width=0.38\linewidth]{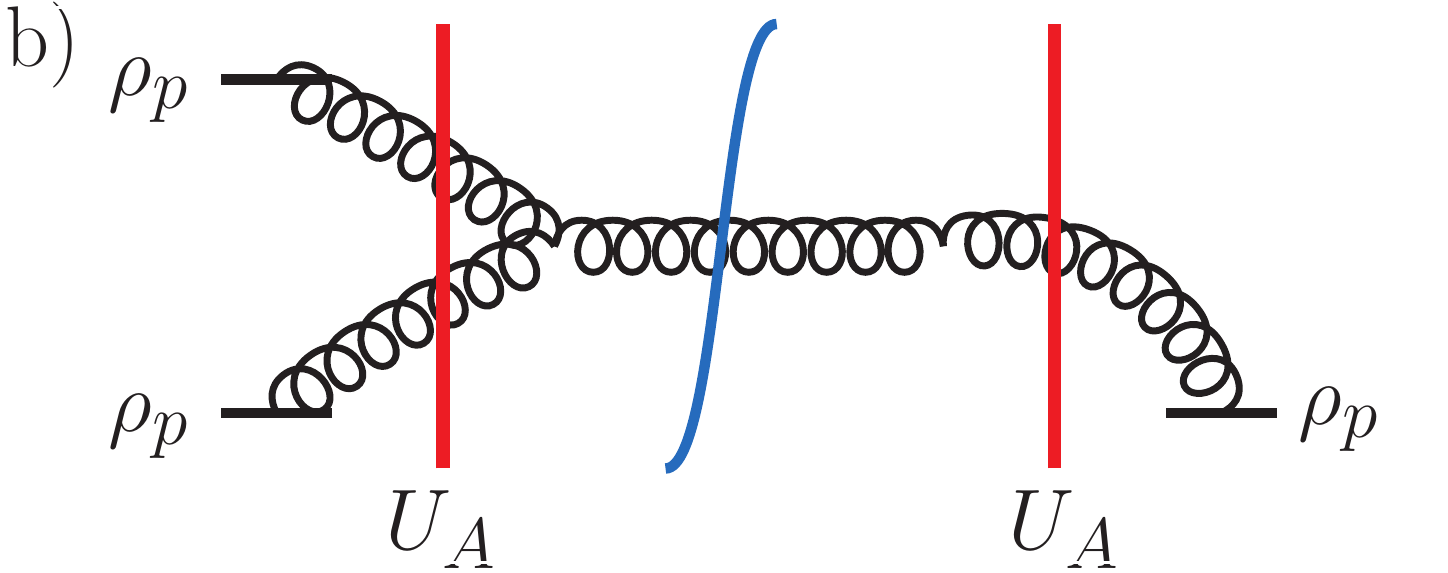}
\caption{Diagrammatic representation of the leading order contributions to the a) P-even and b) P-odd parts of the inclusive gluon production cross section; the straight vertical lines represent multiple scattering in the gluon ``shockwave" field of the target. The curved vertical line denotes the cut separating the amplitude from the complex conjugate amplitude.}
\label{fig:diag}
\end{figure*}

Indeed, as shown explicitly recently~\cite{McLerran:2016snu,Kovner:2016jfp,Kovchegov:2018jun}, the first nontrivial $\rho_p/k_{\perp,p}^2$ correction breaks the accidental parity symmetry and gives a finite contribution to $v_3$. Thus a nonzero value for odd azimuthal anisotropies in the CGC EFT can be understood to be a unique signature of the {\it emerging coherence} of the classical gluon field in the projectile. To quantify these statements, we will follow our previous work and decompose  the single particle inclusive gluon distribution in the dilute-dense CGC EFT into  parity-even and parity-odd contributions~\cite{Mace:2018vwq},
\begin{widetext}
\begin{align}
\label{eqn:dNevenodd}
	%&\frac{dN (\v{k})}{d^2k dy}\Big[\rho_p, \rho_t\Big] =
	%\frac{dN^{\rm even}(\v{k})}{d^2k dy}\Big[\rho_p, \rho_T\Big] +
	%\frac{dN^{\rm odd}(\v{k})}{d^2k dy}\Big[\rho_p, \rho_T\Big]\,, \quad \quad %\notag\\ &
	\frac{dN^{\rm even,\, odd}(\v{k})}{d^2k dy}%\Big[\rho_p, \rho_t\Big]
	=
	\frac12 \left( \frac{dN (\v{k})}{d^2k dy}\,\Big[\rho_p, \rho_t\Big] \pm
	\frac{dN (-\v{k})}{d^2k dy}\Big[\rho_p, \rho_t\Big] \right)\,,
\end{align}
where, employing the analytical results in \cite{Kovchegov:1998bi,Dumitru:2001ux,Blaizot:2004wu,Kovner:2012jm,Kovchegov:2012nd}, one obtains the  
parity-even and parity-odd expressions compactly as~\cite{McLerran:2016snu,Kovchegov:2018jun},
\begin{align}
	\label{even}
	\frac{d N^{\rm even} (\v{k})}{d^2k dy} \Big[\rho_p, \rho_t\Big] &=  \frac{2}{(2\pi)^3}
		\frac{ \delta_{ij} \delta_{lm}  +  \epsilon_{ij} \epsilon_{lm} }{k^2}
		\,\Omega^a_{ij} (\v {k})
		\left[ \Omega^a_{lm} (\v {k}) \right]^\star\,,\\
	\label{odd}
	  \frac{d N^{\rm odd} (\v{k})}{d^2 k dy} \Big[\rho_p, \rho_t  \Big]
    &=
	{ \frac{2}{(2\pi)^3} }
	{\rm Im}
	\Bigg\{
		\frac{g}{{\v{ k}}^2}
		\int \frac{d^2 l}{(2\pi)^2}
				\frac{  {\rm Sign}({\v{k}\times \v{l}}) }{l^2 |\v{k}-\v{l}|^2 }
		f^{abc}
			\Omega^a_{ij} (\v{l})
			\Omega^b_{mn} (\v{k}-\v{l})
			\left[\Omega^{c}_{rp} (\v{k})\right]^\star
		 \\ & \times \quad
		\left[
			\left(
			{\v{ k}}^2 \epsilon^{ij} \epsilon^{mn}
		-\v{l} \cdot (\v{k} - \v{l} )
		(\epsilon^{ij} \epsilon^{mn}+\delta^{ij} \delta^{mn})
		\right) \epsilon^{rp}+
		2 \v{k} \cdot (\v{k}-\v{l}) { \epsilon^{ij} \delta^{mn}} \delta^{rp}
		\right]
	\Bigg\} \, .\notag
\end{align}
\end{widetext}
Here
\begin{equation}
	\Omega_{ij}^a(\v{k}) =
	g \int \frac{d^2 p }{(2\pi)^2}
	\frac{p_{i} (k-p)_{j} }{p^2}
	\rho^b_p(\v{p}) U_{ab} (\v{k}-\v{p})\,,
	\label{Eq:Omega}
\end{equation}
and $\epsilon_{ij} (\delta_{ij})$ denotes the Levi-Civita symbol (Kronecker delta). The adjoint Wilson line $U_{ab}$ is a functional of the target charge density and is the two-dimensional Fourier transform of its coordinate space counterpart:
\begin{equation}
\tilde U(\v{x}) = {\cal P} \exp \left( i g^2 \int dx^+ \frac{1}{\v{\nabla}^2}{\tilde \rho}_t^a(x^+, \v{x}) T_a \right). 
	\label{Eq:U}
\end{equation}
If we compare the even and odd contributions in Eqs.~\ref{even} and \ref{odd} respectively, one observes that the latter is suppressed in the CGC EFT power counting by $\alpha_S \rho_p$, where $\alpha_S = g^2/4\pi$ is the QCD coupling.
This is also apparent from the diagrammatic representation of the inclusive single particle distribution depicted in Fig.~\ref{fig:diag}.

Defining the harmonics of the even and odd single particle azimuthal momentum anisotropies, for a fixed configuration of $\rho_p$ and $\rho_t$, respectively as
\begin{equation}
	Q_{2n} \left[\rho_p,\rho_t\right] = \frac{ \int_{p_1}^{p_2}   k_\perp dk_\perp\frac{d\phi}{2\pi} e^{i 2 n \phi} \frac{d N^{\rm even} (\v{k})}{d^2k dy} \Big[\rho_p, \rho_t\Big]
  }     {   \int_{p_1}^{p_2}   k_\perp dk_\perp \frac{d\phi}{2\pi} \frac{d N^{\rm even} (\v{k})}{d^2k dy} \Big[\rho_p, \rho_t\Big] } \,,
\end{equation}
\begin{equation}
	Q_{2n+1} \left[\rho_p,\rho_t\right]  = \frac{ \int_{p_1}^{p_2}   k_\perp dk_\perp\frac{d\phi}{2\pi} e^{i (2 n+1) \phi} \frac{d N^{\rm odd} (\v{k})}{d^2k dy} \Big[\rho_p, \rho_t\Big]
  }     {   \int_{p_1}^{p_2}   k_\perp dk_\perp \frac{d\phi}{2\pi} \frac{d N^{\rm even} (\v{k})}{d^2k dy} \Big[\rho_p, \rho_t\Big] } \,,
\end{equation}
the physical two-particle anisotropy coefficients can be simply expressed as~\footnote{In the following, for notational simplicity, we will not explicitly write the limits of the momentum integration arguments.}
\begin{eqnarray}
	v_n^2\{2\} (N_{\rm ch})%(p_\perp) %(|\v{k}_1|, |\v{k}_2|)
	&=   \int {\cal D} \rho_p {\cal D} \rho_t\  W[\rho_p] \ W[\rho_t] 
	  \\
	  &\quad \times
	|Q_n \left[\rho_p,\rho_t\right] |^2 %(|\v{k}_2|)
  \delta \left( 	\frac{d N }{dy} \Big[\rho_p, \rho_t\Big] - N_{\rm ch}  \right) \notag
	\,.
\label{eqn:vn-formula}
\end{eqnarray}
The weight functionals representing the distribution of color sources,
\begin{equation}
	W[\tilde\rho_{p,t}] =  {\cal N} \exp\left[  -\int dx^{-,+} d^2 x \frac{\tilde\rho_{p,t}^a(x^{-,+} ,\v{x})  \tilde\rho_{p,t}^a  (x^{-,+} ,\v{x})  }{2 \mu^2_{p,t}(\v{x}) }\right]\,,
\end{equation}
have the McLerran-Venugopalan (MV) model~\cite{McLerran:1993ni,McLerran:1993ka} form, where ${\cal N}$ is a normalization factor. However, unlike the MV model, $\mu^2_{p,t}(\v{x})$, the color charge squared per unit area, is spatially dependent here due to i) the renormalization group (RG) evolution of the color sources to small Bjorken $x$ for the case of  $\mu^2_p$~\cite{Dumitru:2011vk,Dusling:2009ni,Schenke:2012wb}, and ii) both RG evolution and fluctuations in the nucleon positions in the target for $\mu^2_t$.

\begin{figure*}
\includegraphics[width=0.48\linewidth]{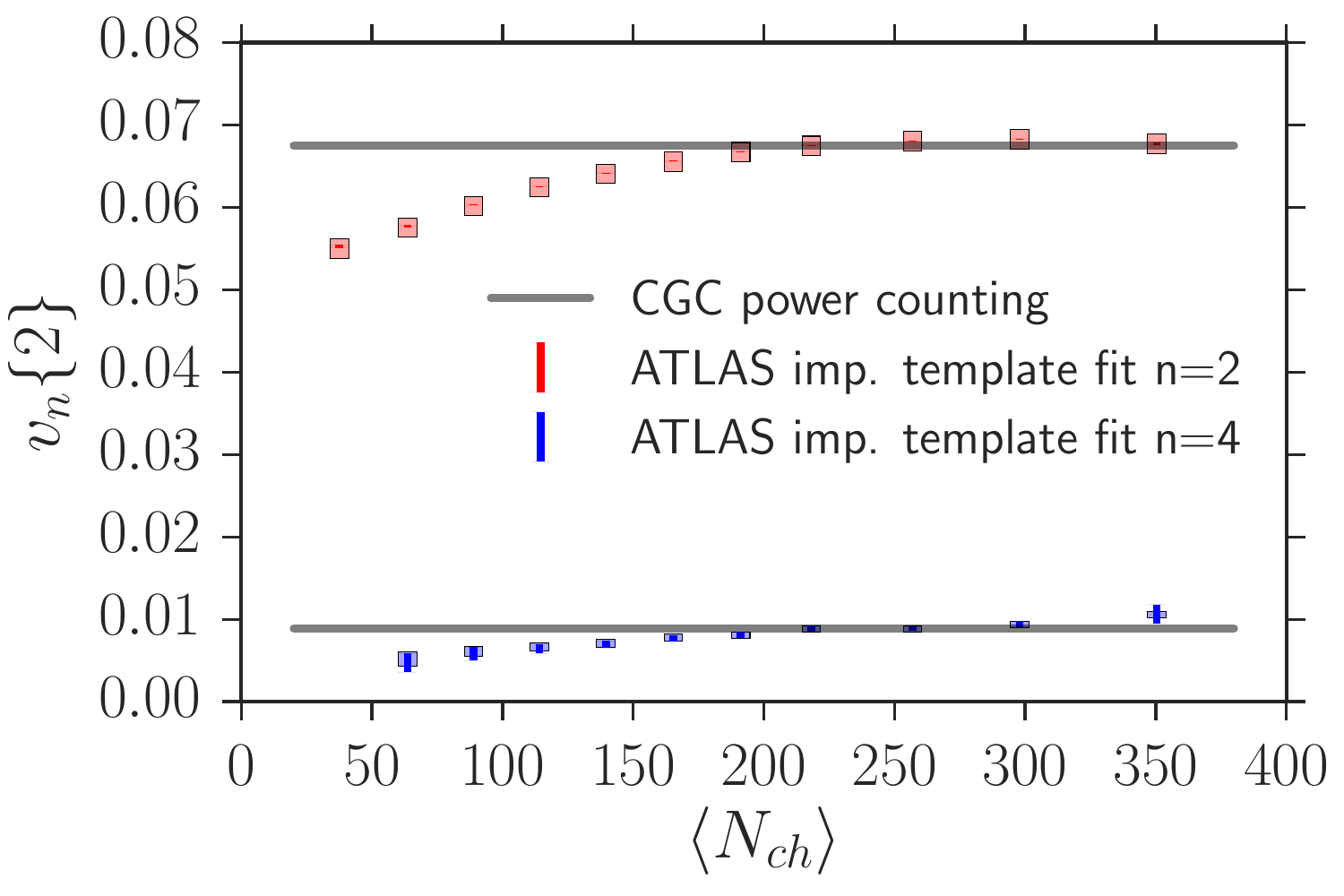}\hspace{0.1cm}
\includegraphics[width=0.48\linewidth]{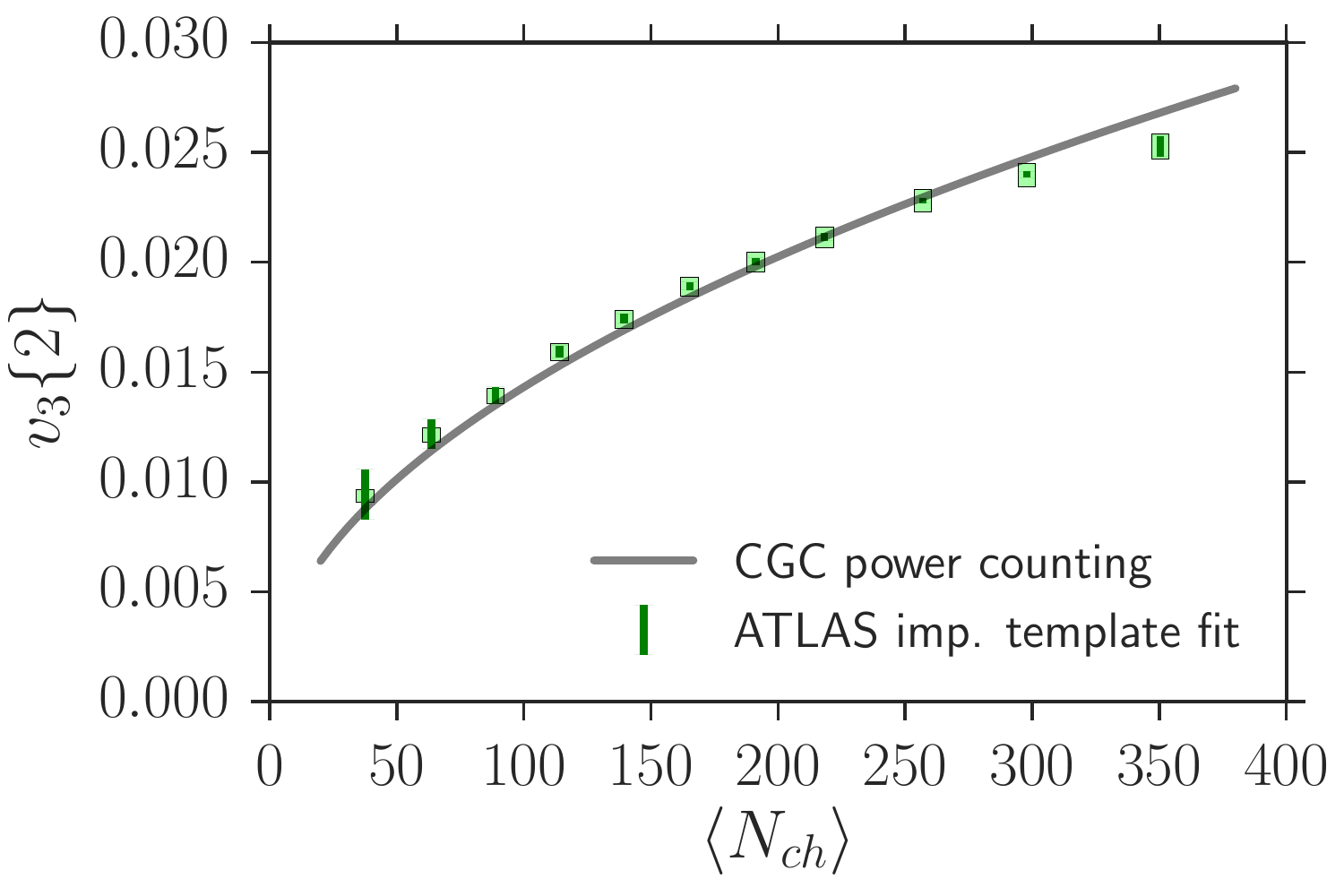}
\caption{Comparison of even and odd harmonics measured in p+Pb collisions from ATLAS improved template fit~\cite{ATLAS-CONF-2018-012} with the $N_{\rm ch}$ scaling expected in the dilute-dense CGC EFT. }
\label{fig:lhc_comp}
\end{figure*}

%
%The multiplicity distribution in high energy p+A collisions is determined primarily by the color fluctuations in the projectile wave function~\cite{Liou:2016mfr,Kovner:2018azs} encoded in $\rho_p$. Additionally, there are fluctuations in the target wavefunction, including Glauber $N_{\rm part}$ fluctuations; in the dilute-dense regime, target fluctuations contribute to the particle multiplicity {\it logarithmically}~\cite{Dumitru:2001ux}. These can therefore, to first approximation, be neglected, a statement that becomes increasingly robust with inreasing $N_{\rm ch}$.   
%%We will return to this point later with an illustrative example. 

To examine the multiplicity dependence of $v_n$, let us first rescale $\rho_p \to c \rho_p$, which gives 
\begin{eqnarray}
	\Omega(\v{k}) \to c \ 	\Omega(\v{k}) \ ,
\end{eqnarray}
in Eq.~\eqref{Eq:Omega}. 
This property is obvious physically because  $\Omega(\v{k})$ represents the covariant gauge classical field 
of the projectile color rotated by the Wilson line of the target.  Hence the ``event'' multiplicity, for a fixed configuration of $\rho_p$, $\rho_t$, 
transforms as  
\begin{eqnarray}
\label{eqn:dndy_scal}
  \frac{d N }{dy} \Big[\rho_p, \rho_t\Big] \to c^2 \frac{d N }{dy} \Big[\rho_p, \rho_t\Big]
  + {\cal O} (c^3)\,.
\end{eqnarray}
The order $c^3$ contribution to the multiplicity is  P-odd and will vanish after performing the ensemble average. The first nontrivial 
correction is thus of order $c^4$ and can be interpreted as the first saturation correction from the proton to single inclusive gluon  production~\cite{Chirilli:2015tea}. 
%In this discussion we consider the leading order contributions to physical observables 
%only and neglect higher order corrections. We comment on the higher density corrections at the and.   

%\begin{widetext}
The even ``single particle'' harmonic is invariant under rescaling 
\begin{eqnarray}
\label{eqn:phieven_scal}
	Q_{2n} \left[\rho_p,\rho_t\right] \to c^0 \ Q_{2n} \left[\rho_p,\rho_t\right]\, ,
\end{eqnarray}
because both numerator and denominator scale identically.  In contrast, the P-odd contribution appears in the numerator of the odd single particle harmonic, while the normalization in the denominator is dominated by the P-even leading order piece. One therefore obtains,
\begin{eqnarray}
\label{eqn:phiodd_scal}
	Q_{2n+1} \left[\rho_p,\rho_t\right]   \to c \ Q_{2n+1} \left[\rho_p,\rho_t\right]\,.
\end{eqnarray}
%\end{widetext}
Hence
% since particle production in the dilute-dense CGC EFT is driven by fluctuations in the
%projectile color charge density,  
these scaling relations Eqs.~\eqref{eqn:dndy_scal}-\eqref{eqn:phiodd_scal} allow us to establish that 
\begin{eqnarray}
  	Q_{2n} \left[\rho_p,\rho_t\right] & \propto \left( \frac{d N }{dy} \Big[\rho_p, \rho_t\Big] \right)^0, \\
	Q_{2n+1} \left[\rho_p,\rho_t\right]& \propto \sqrt{  \frac{d N }{dy} \Big[\rho_p, \rho_t\Big]}  \,,
\end{eqnarray}
and therefore, 
\begin{eqnarray}
  v_{2n}\{2\} \propto N_{\rm ch}^0,  \quad \quad
    v_{2n+1}\{2\} \propto N^{1/2}_{\rm ch}\,.
\end{eqnarray}
This argument is insufficient to fix the coefficients of proportionality. These can however be fixed by data on even and odd harmonics at a given $N_{\rm ch}$.  In order to do this, we choose $N_{\rm ch}=218$ for all $v_n$. We plot the results of this scaling for the azimuthal Fourier harmonics $v_2\{2\}$, $v_3\{2\}$, $v_4\{2\}$ versus p+Pb ATLAS improved template fit data~\cite{ATLAS-CONF-2018-012} in Fig.~\ref{fig:lhc_comp}. We have checked that the results shown are relatively insensitive to the value of $N_{\rm ch}$ used to extract the proportionality coefficient for each harmonic.  Note that since this template fit method aims to extract the long-range ridge correlations by cleanly separating them from dijet contribution, it is ideal for theory to data comparisons.  

Remarkably, we find that the anticipated scaling of  $v_{2,3,4}$  with $N_{\rm ch}$, obtained from the CGC power counting, is in excellent agreement with the ATLAS data. It is unclear at present what the corresponding qualitative expectations are in kinetic and hydrodynamic models for p+A collisions. 
Studies which include a dense-dense CGC initial state, followed by hydrodynamic evolution~\cite{Mantysaari:2017cni}, also exhibit relative independence on ${N_{\rm ch}}$ for $v_2$. However for $v_3$, they do not see the $\sqrt{N_{\rm ch}}$
dependence but a much flatter behavior. This flatter behavior is also anticipated in the CGC EFT power counting for large $N_{\rm ch}$ because $\rho_p$ from the projectile is promoted to a Wilson line $U$ in the dense-dense limit. It is therefore important to quantify whether the results of \cite{Mantysaari:2017cni} are due to 
the dense-dense IP-Glasma initial dynamics~\cite{Schenke:2012wb,Schenke:2013dpa,Schenke:2015aqa,Schenke:2016lrs} or whether the later (and relatively short) hydrodynamic expansion is also essential in p+Pb collisions.

\begin{figure}
\includegraphics[width=0.96\linewidth]{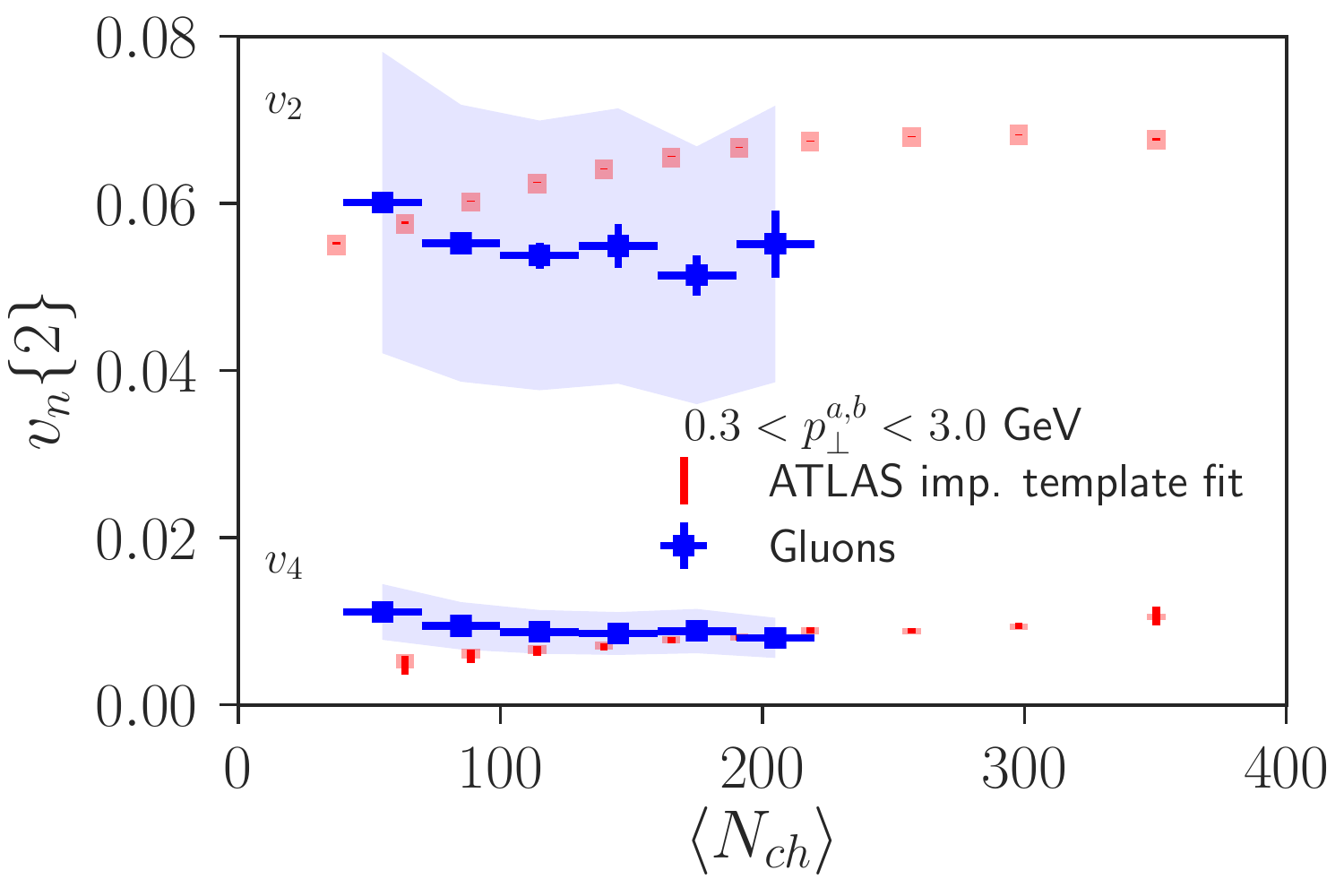}
\caption{Comparison of even and odd harmonics measured in p+Pb collisions from ATLAS improved template fit~\cite{ATLAS-CONF-2018-012} with the dilute-dense CGC EFT. A 30\% systematic uncertainty band is shown.}
\label{fig:veven_numerics}
\end{figure}

We will now employ the numerical realization of the dilute-dense framework we developed in \cite{Mace:2018vwq} to quantitatively verify whether they corroborate our  simple power counting estimates. In Fig.~\ref{fig:veven_numerics}, we show our results for $v_2$ and $v_4$. The parameters in our study are as follows \footnote{See also 
\cite{Mace:2018vwq} and \cite{Schenke:2013dpa} for more details on the parameters entering solutions of the QCD Yang-Mills equations.}. We take the ratio of the saturation scale to the color charge scale to be $Q_s/g^2\mu=0.5$; additionally we take the fluctuations of the log of this ratio to be $\sigma=0.5$ -- this quantity is important to account for the large color charge fluctuations in rare events.~\cite{McLerran:2015qxa}. Additionally, we employ a 
regulator mass of $m=0.3$ GeV for the gauge fields
which enters the Poisson equation relating the gauge fields in the projectile and target to their respective color charge densities. As in our previous study~\cite{Mace:2018vwq}, these parameters are determined by minimizing the deviation from the charged particle multiplicity distribution measured by the ATLAS collaboration~\cite{Aaboud:2016jnr}. Further, following the lattice prescriptions of \cite{Lappi:2007ku}, we take transverse lattices of size $N=1024$ with lattice  spacing $a=0.0625$ fm, and $N_y=100$ rapidity slices in the coordinate $x^-$; we have verified that the continuum limit is obtained for this parameter set.

\begin{figure}
\includegraphics[width=0.96\linewidth]{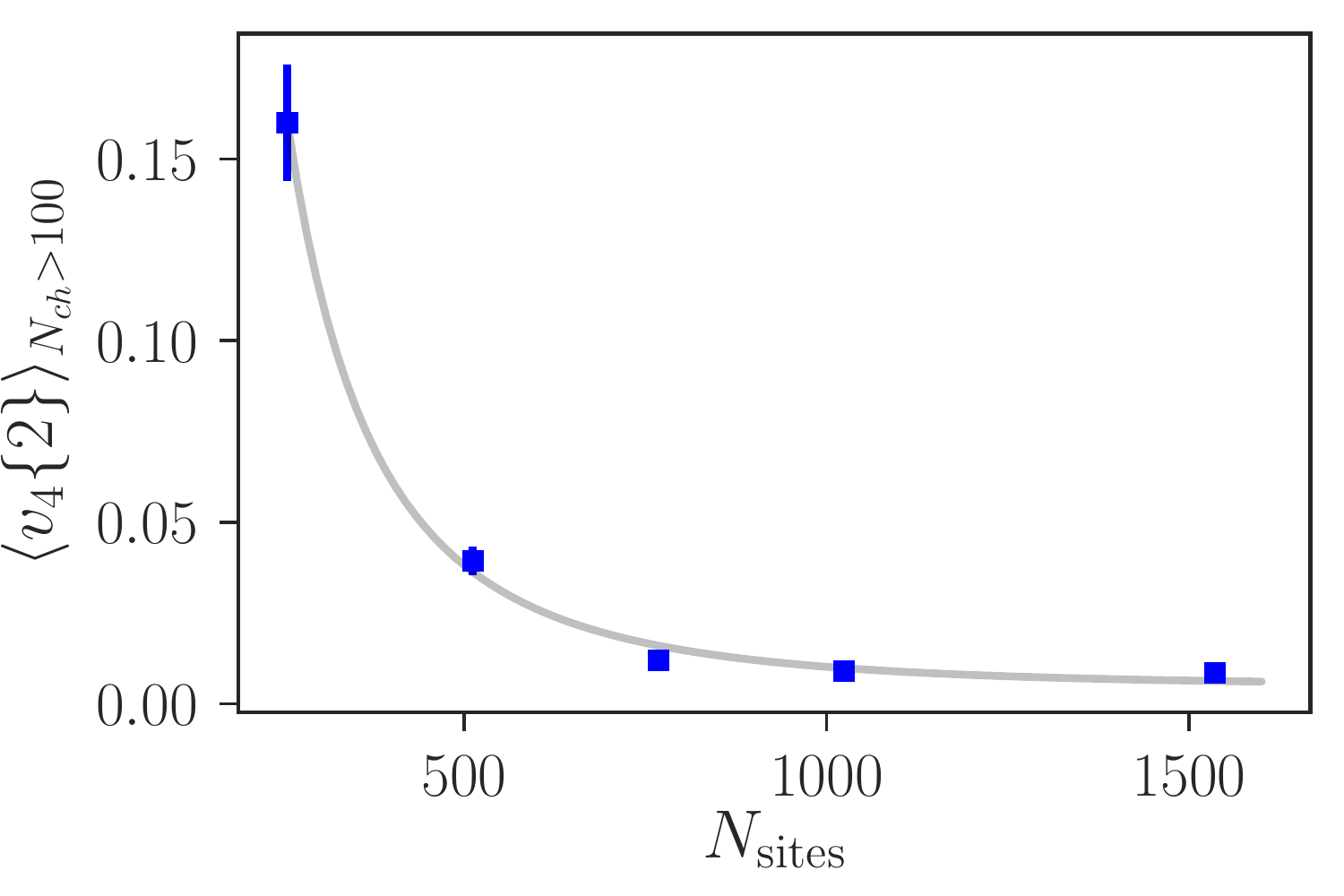}
\caption{Lattice size dependence (for a fixed spacing of $a=0.0625$ fm) of the average $v_4\{2\}$ for $N_{\rm ch}\geq 100$. Excellent convergence is observed for $N\geq 768$ lattice sites.}
\label{fig:v4continuum}
\end{figure}

The extraction of $v_4$ is especially challenging because very fine lattices are required to extract a robust fourth harmonic. Our results for the convergence of $v_4$ with increasing lattice size are shown in Fig.~\ref{fig:v4continuum}. These results also suggest that a robust extraction of $v_4$ in the numerically more intensive dense-dense IP-Glasma~\cite{Schenke:2015aqa} framework for small systems is 
very challenging
 %likely impossible
with current resources. The numerical extraction of $v_3$ in the dilute-dense CGC EFT as a function of $N_{\rm ch}$  is even more challenging than that for $v_{2,4}$. This is evident from Eq.~\eqref{odd} because firstly, one more $\rho_p$ and $U$ distribution have to be sampled; further, one needs to extract the imaginary piece of this expression. For $v_{2,4}$ one needs approximately $10^4$ color charge configurations; for $v_3$ we will require at least $10^6$ configurations. This is challenging and outside the scope of the present work. %--we intend to pursue it in future. 

Our results for $v_{2,4}$ as a function of $N_{\rm ch}$ are shown in Fig.~\ref{fig:veven_numerics}.  
Unfortunately, the computational effort required limits our ability to go to larger $N_{\rm ch}$. Within the systematic uncertainties enumerated below, the agreement of data with the theory is very good. These uncertainties include the dependence on the regulator mass $m$ and on the ratio $Q_s/g^2\mu$ and its fluctuations. Our studies varying these quantities suggest that the uncertainties are of order 30\% percent. This is represented by the shaded band around our results for the values quoted above for these parameters. In addition, there are uncertainties from running coupling corrections~\cite{Schenke:2015aqa} and from hadronization of gluons~\cite{Schenke:2016lrs}. These uncertainties are somewhat mitigated by the fact that the $v_n$'s are ratios of weighted multiplicities.

It is natural to ask if the correlations in p+Pb collisions have the same origin as those measured in peripheral Pb+Pb
collisions at the same $N_{\rm ch}$.
The CMS collaboration \cite{Chatrchyan:2013nka} observed that the two-particle $v_3$ in p+Pb and Pb+Pb collisions are nearly identical for the same value of $N_{\rm ch}$ while $v_2$ is larger in the latter system. 
The IP-Glasma+MUSIC model~\cite{Gale:2012rq} (wherein the dense-dense CGC EFT IP-Glasma initial conditions
are combined with hydrodynamic flow~\cite{Schenke:2010rr}) which does a good job of describing Pb+Pb data for central collisions out to at least 60\% centrality,
does a poor job of describing the p+Pb $v_{2,3}$ data~\cite{Schenke:2014zha}. 
%
%In \cite{Schenke:2014zha}, it was shown that the IP-Glasma+MUSIC model~\cite{Gale:2012rq} (wherein the dense-dense CGC EFT IP-Glasma initial conditions are combined with hydrodynamic flow~\cite{Schenke:2010rr}) does a poor job of describing the p+Pb $v_{2,3}$ data.
%
It was recently shown~\cite{Mantysaari:2017cni} that including spatial fluctuations of color charge distributions~\footnote{These fluctuations have been argued to be necessary to describe HERA data on incoherent exclusive $J/\Psi$ production~\cite{Mantysaari:2016ykx,Mantysaari:2016jaz}.} in the model has a big effect leading to good agreement with the p+Pb data. 
 The impact of these studies for peripheral Pb+Pb collisions remains to be quantified. 
The results of both \cite{Schenke:2014zha} and \cite{Mantysaari:2017cni} rely on hydrodynamic response to an initial spatial geometry and is qualitatively different from our effect which results from initial state momentum anisotropies alone. 

Will our dilute-dense framework suffice to describe $v_{2,3}$ in peripheral Pb+Pb collisions?  Qualitatively, we anticipate that the $N_{\rm ch}$ dependence of $v_{2,3}$ would be identical for p+Pb and Pb+Pb for partons in the projectile coherently interacting with multiple localized color domains of size $1/Q_s$~\footnote{Interestingly, similar arguments on universal $N_{\rm ch}$ scaling have been advanced in a kinetic theory picture~\cite{Basar:2013hea}.}. For the values of $N_{\rm ch}$ studied here, $v_2$ is approximately 25\% greater in Pb+Pb than in p+Pb. We plan to examine whether this difference is captured in quantitative studies for Pb+Pb in the dilute-dense and dense-dense  frameworks~\cite{MSTV-paperIII}. Alternatively, because the size of the Pb+Pb system is larger, and rescattering is more likely to  occur~\cite{Greif:2017bnr,Kurkela:2018vqr}, the breakdown of scaling may affect $v_2$ sooner than it does for $v_3$. Both of these scenarios for peripheral Pb+Pb collisions, as well as the relative role of initial momentum anisotropy versus enhanced geometry response from shape fluctuations in p+Pb collisions, can be quantified within  the CGC EFT framework itself and will be reported in the near future~\cite{Bjoern-Chun-Mark-Prithwish-Vladi-Raju}.

We thank Peter Braun-Munzinger, Adam Bzdak, Brian Cole, Adrian Dumitru,
Jiangyong Jia, Larry McLerran, Soumya Mohapatra,  Wei Li, Volker Koch, Jean-Yves Ollitrault,
Bj\"{o}rn Schenke, S\"{o}ren Schlichting, Juergen Schukraft, Chun Shen and
Derek Teaney for useful discussions. This material is based on work supported
by the U.S. Department of
Energy, Office of Science, Office of Nuclear Physics, under Contracts No.
DE-SC0012704 (M.M.,P.T.,R.V.) and DE-FG02-88ER40388 (M.M.). M.M. would also
like to thank the BEST Collaboration for support. This research used resources
of the National Energy Research Scientific Computing Center, a DOE Office of
Science User Facility supported by the Office of Science of the U.S. Department
of Energy under Contract No. DE-AC02-05CH11231.

%\bibliography{bibl.bib}
\bibliography{bibl_insp.bib}

\end{document}